\documentclass{wbl} 

\usepackage{makeidx}
\makeindex
\usepackage{epsfig}

\catcode`\.=\active\def.{\char'56\allowbreak}\catcode`\.=12
\catcode`\/=\active\def/{\char'57\allowbreak}\catcode`\/=12
\catcode`\:=\active\def:{\char'72\allowbreak}\catcode`\:=12  
\catcode`\@=\active\def@{\char'100\allowbreak}\catcode`\@=12 
\catcode`\-=\active\def-{\char'55\allowbreak}\catcode`\-=12  
\def\URL{\bgroup\catcode`\.=\active\catcode`\/=\active\catcode`\:=\active\catcode`\@=\active\catcode`\-=\active\def~{\char126}\tt\URLaux}
\def\URLaux#1{\href{#1}{#1}\egroup}

\newlength{\halfwidth}
\newlength{\rocpicoff} 
\newlength{\temp}
\newlength{\tempa}

\begin{document}

\setcounter{chapter}{10} 
\renewcommand{\thepage}{\arabic{page}}


\articletitle{The Distribution of\mbox{ }\\
Reversible Functions is Normal}

\author{W. B. Langdon}

\email{W.Langdon@cs.ucl.ac.uk}

\affil{Computer Science,
        University College, London,\\
        Gower Street,
        London, WC1E 6BT, UK
}
\date{13 March 2003  $Revision: 1.36 $}

\begin{abstract}
The distribution of reversible programs tends to a limit
as their size increases.
For problems with a Hamming distance fitness function
the limiting distribution is 
binomial with an exponentially small chance 
(but non~zero)
chance of perfect solution.
Sufficiently good reversible circuits are more common.
Expected RMS error is also calculated.
Random unitary matrices may suggest possible extension to quantum computing.
Using the genetic programming (GP) benchmark,
the six multiplexor,
circuits of Toffoli gates are shown to give a fitness landscape
amenable to evolutionary search.
Minimal CCNOT solutions to
the six multiplexer are found
but larger circuits are more evolvable.
\end{abstract}



\chapbblname{reversible}
\normallatexbib
\bibliographystyle{kapalike}

 
\index{Reversible computing|(}

\section{Introduction}
\label{intro}

\index{Hamming distance fitness}

We shall show the fitness of classical reversible computing programs
\cite{bennett:1985:sciam}
(where fitness is given by Hamming distance from an ideal answer)
is Normally distributed.
If the score is normalised so that the maximum score (fitness)
is 1 and the minimum is 0,
then the
mean is 0.5 and the standard deviation is
$1/2\ m^{-1/2} 2^{-\frac{n}{2}}$.
(Where $n$ is the number of input bits and $m$ is the number of output
bits.)

Almost all genetic programming has used traditional computing
instructions,
such as add, subtract, multiple, or, and.
These instruction sets are not reversible.
I.e., in general, it is impossible given a program and its output,
to unambiguously reconstruct the program's input.
This is because most of the primitive operations themselves are irreversible.
However genetic programming can evolve reversible programs
composed of reversible primitives.

\index{Reversible gate arrays}
\index{Reversible circuits}
\index{Quantum circuits}
\index{Low heat dissipation}
\index{Information loss}

A number of reversible gates have been proposed
\cite{ICALP::Toffoli1980,FreTof82}
which can be connected in a linear sequence to give a reversible gate
array,
which we will treat as a reversible computer program.
At present the driving force behind the interest in reversible
computing is the hope that reversible gates can be implemented as
quantum gates, leading to quantum coherent circuits and quantum computing.
Reversible computing has also been proposed for safety critical
applications
and for low power consumption or low heat dissipation.

In the absence of counter measures,
most traditional computer programs degrade information.
I.e.\
knowledge about their inputs is progressively lost as they are
executed.
This means,
most programs
produce the same output regardless of their input
\cite{langdon:2002:foga}.
Suppose a program has $n$ input bits 
and $m$ output bits,
there are
$2^{m 2^{n}}$ possible functions
it could implement.
However a long program is almost certain to implement one of the 
$2^{m}$ constants.
That is, the fraction of functions actually implemented is tiny
as the programs get longer
and worse,
the fraction of interesting functions tends to zero.
This is due to the inherent irreversibility of traditional computing
primitives.

In the next section we describe reversible computing
in more detail.
In reversible computing there is also a distribution of functions
which programs tend to as they get longer.
Instead of it being dominated by constants,
every {\em reversible} function is equally likely
(cf.\
Section~\ref{sec:proof}).
Convergence to this limit is tested 
in Section~\ref{sec:random}
on the Boolean
6~Multiplexor problem (described in Section~\ref{sec:6mux}).
Section~\ref{sec:random} shows convergence to the large program limit
can be rapid.
Despite the low density of solutions,
evolutionary search is effective at finding them
(Section~\ref{sec:mutation}).
Initial results suggest,
CCNOT
and
traditional computing primitives
are similarly amenable to evolutionary search.

\section{Reversible Computing Circuits}
\label{sec:rev}
A reversible computer can be treated as an array of parallel wires
leading from the inputs and constants to the outputs and garbage.
In normal operation the garbage outputs are treated as rubbish and
\index{Garbage outputs}%
\index{Rubbish outputs}%
discarded by the end of the program.

\begin{figure}
\centerline{\includegraphics{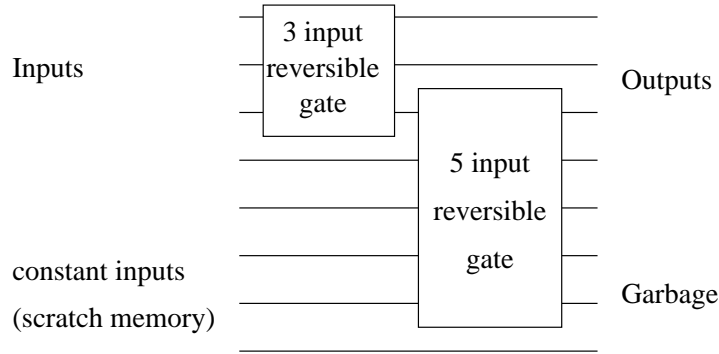}} 
\caption{\label{fig:computer}
Schematic of reversible computer.
In normal operation data flows from left to right.
However when reversed, garbage and outputs are fed into the circuit
from the right
to give the
original inputs to the program (plus constants).
}
\end{figure}

\begin{figure}
\includegraphics{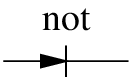}
\hfill
\includegraphics{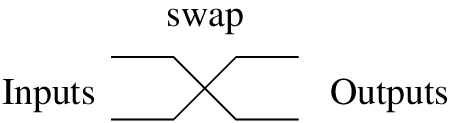}
\hfill
\includegraphics{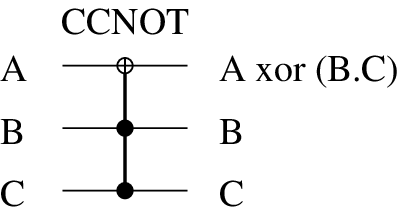}
\\
\caption{\label{fig:gates}
Examples of reversible gates.
The CCNOT (Toffoli) gate passes all of its inputs directly to
its outputs, 
unless all of the control lines (B and C) are true,
in which case it inverts its main data line
(A).
(The control inputs of a CCNOT gate (B and C) 
can be connected to the same wire but
the controlled wire (A) cannot be directly connected to either control.)
CCNOT is complete, in the sense all boolean circuits can be
constructed from it.
\index{Not (reversible)}
\index{Swap}
\index{CCNOT}
\index{Controlled-controlled not|see{CCNOT}}
\index{Toffoli gate|see{CCNOT}}
}
\end{figure}

Connected across the wires are reversible gates.
Each gate has as many inputs as it has outputs.
The gates are reversible, in the sense that it is possible to
unambiguously identify their inputs given their outputs
(see Figure~\ref{fig:computer}).
The simplest reversible gate is the identity,
i.e.\
a direct connection from input to output.
Also NOT is reversible,
since given its output we know what its input must have been.
Similarly a gate which swaps its inputs is also reversible
(see Figure~\ref{fig:gates}).

\index{CCNOT|(}
We will mainly be concerned with the
controlled-controlled not
(CCNOT, Toffoli) gate.
Unless all of its control lines are true,
the CCNOT gate passes all of its inputs directly to
its outputs.
However if they are all set,
CCNOT inverts the controlled line.
CCNOT is complete in the sense,
given sufficient CCNOT gates and additional
constant inputs and rubbish outputs,
a reversible circuit 
equivalent to any Boolean function from inputs to outputs
(excluding constants and rubbish)
can be constructed.
Since CCNOT can invert~1,
the additional inputs can all be~1.
Note a single CCNOT gate 
(plus a constant zero,
e.g.\ 
provided by using another CCNOT gate to invert a one)
can implement the identify function.
In C code 
\protect\verb!data[A] = (data[C] & data[B]) ^ data[A]; !
\index{CCNOT|)}

\index{Permutation|(}
While it is not necessary 
for the number of lines set to true, to remain constant across the circuit from
left to right,
a reversible computer must implement a permutation.
To explain what we mean by this,
consider the left hand side (of $N$~wires) as an $N$~bit number.
There are up to $2^{N}$ possible left hand patterns.
Similarly there are up to $2^{N}$ possible right hand patterns.
The computer provides a mapping from left hand number to right hand number.
For the mapping to be reversible,
its range and domain must be the same size
and 
a number can only appear once on the right hand side (range),
i.e.\
the mapping must be a permutation.
If all $2^{N}$ possible numbers are used,
only $2^{N}!$ of the 
$2^{N 2^{N}}$ possible mappings are reversible.
For large $N$, 
this means only
about 1 in $e^{2^{N}}$ mappings 
are permutations
and hence are reversible.
\index{Permutation|)}

\index{Energy consumption limit|(}
The computation remains reversible up until the garbage bits are
discarded.
It is at this point that 
information is lost.
It is the deletion of information which means the
computation must consume energy and release
it as heat.
By carefully controlling the deletion of these rubbish bits,
it has been suggested that reversible computers will require less
energy than irreversible computers.
Present day circuits do not approach the lower bound on energy
consumption suggested by their irreversibility.
I.e.\
they require much more energy to operate gates, drive
connecting wires, etc.\
than 
the theoretical bound on energy consumption due to 
information lost as they run.
\index{Energy consumption limit|)}

\index{Cooling computers|(}
\index{Heating circuits|(}
However in the near term,
energy consumption is interesting both for ultra-low power
consumption,
e.g.\
solar powered computing,
and also because the energy released inside the computing circuit has
to be removed as heat.
The only way heat is removed at present is by making the centre of the
circuit hotter and allowing heat to diffuse down the temperature
gradient to the cooler boundaries of the circuit.
(Active heat pumps within the circuit have been considered.
Electronic refrigerators
could be based upon the Peltier electro thermodynamic effect).
\index{Peltier effect}%
Even today cooling circuits is a limiting factor on their operation.
Increasing circuit clock speeds,
despite continued reduction in circuit size,
mean heat removal will be an increasing concern.
\index{Cooling computers|)}
\index{Heating circuits|)}

\index{Safety critical software!reversible computing|(}
\cite{Bishop97} describes a single channel
reversible system for a safety critical
control application. By running the system forwards and then backwards
and comparing the original inputs with those returned by traversing the
system twice,
he demonstrated the system was able to detect test errors
injected into the system during its operation.
(High reliability systems often use comparison between multiple
channels to detect errors.)
\index{Safety critical software!reversible computing|)}

\section{Distribution of Large Reversible Circuits}
\label{sec:proof}

\index{Permutation!uniform limit|(}
As with a complete reversible circuit,
the action of a single reversible gate across $N$ wires can be treated
as a permutation mapping  one $N$~bit number to another.
Following 
\cite{langdon:fogp},\cite{langdon:2002:crlp},\cite{langdon:2002:foga},
we can treat the sequence of permutations from the start of the circuit to
its end as a sequence of state transitions.
The state being the current permutation.
We restrict ourselves to just those permutations which can be implemented,
i.e.\
states that can be reached.
Each gate changes the current permutation (state) to the next.
We can describe the action of a gate by a square matrix of
zeros and ones.
Each row contains exactly one one.
The position of the one indicates the permutation
on the output of the gate
for each permutation on the input side of the gate.
\index{Stochastic matrix|(}%
(Note the matrix is row stochastic).
Now each gate is reversible.
I.e.\
given a permutation on its output side,
there can only be one permutation on its input side.
This means each column of the matrix also contains exactly one one.
(I.e.\
the matrix is column as well as row stochastic,
i.e.\
it is double stochastic).
We will have multiple ways of connecting our gates 
or even multiple types of gate,
however
each matrix will be double stochastic and therefore so too will be the 
average matrix.
Since we only consider implementable permutations,
the average matrix is fully connected.
If a single gate can implement the identity function,
the matrix must have a non~zero diagonal element.
This suppresses cycling in the limit.
If we choose gates at random,
the sequence of permutations is also random.
Since the next permutation depends only on the current permutation
and the gate,
the sequence of permutations is a Markov process.
\index{Markov processes}%
The Markov transition matrix is the average of each of the gate matrices,
which is fully connected, acyclic and double stochastic.
\index{Stochastic matrix|)}%
This means as the number of randomly chosen
gates increases 
each of the Markov states becomes equally likely
\cite{feller:1957:ipta}. 
I.e.\
in the limit of large circuits each possible permutation is equally likely.
\index{Permutation!uniform limit|)}

When there are many randomly connected gates
and 
the total number of lines $N$ is large,
not only is each possible permutation equally likely but
(since our reversible gates are complete)
all $m$~bit output patterns are possible.
Further we will assume that
we can treat each output bit as being 
equally likely to be on as off and
almost independent of the
others.
\index{Hamming distance fitness}%
If fitness $f$ is defined by running the program on every input 
(i.e.\
running it $2^{n}$ times)
and summing the number of output bits that match a target
($f=$ Hamming distance)
then $f$
follows the Binomial distribution
$2^{-m2^{n}}C_{f}^{m2^{n}}$.
This means most programs have a fitness near the average $\frac{1}{2}m2^{n}$
and the chance of finding a solution is $2^{-m2^{n}}$.
While small, this is finite, whereas with irreversible gates
(and no write protection of inputs)
almost all programs do not solve any non~trivial problem
\cite{langdon:2002:foga}.

\index{Root mean squared (RMS) fitness|(}
We can also use the known uniform distribution to calculate
\linebreak[4]
the expected RMS error.
Suppose $T$ of the $2^{n}$ possible fitness 
\linebreak[4]
cases are run, the expected
average squared error is
\linebreak[4]
$\overline{\rm RMS}=$
$\frac{1}{k}\sum_{i=1}^{k}
\sqrt{\frac{1}{T}\sum_{t=1}^{T}|P_i({\rm input}_t+2^{N}-2^{n})\bmod 2^{m}-A_t)|^{2}}$.
Where 
$k$ is the total number of permutations,
${\rm input}_t$ is the input for the $t^{th}$ test,
$A_t =$ required answer
and $P_i(x)$ is the $i^{th}$ permutation of $x$.
As \mbox{$2^{m} \ll k$} we can approximate the average behaviour of
\mbox{$P_i(t+2^{N}-2^{n})\bmod 2^{m}$}
over $k$ cases
by
$i$
over $2^{m}$ cases.
So
the expected
average squared error is
$\overline{\rm RMS}=
\frac{1}{2^{m}}\sum_{i=0}^{2^{m}-1}
\sqrt{\frac{1}{T}\sum_{t=1}^{T}|i-A_t|^{2}}$.
If only a few tests with small values are run
(i.e.\
$T\ll 2^{m}$ and $A_t\ll 2^{m}$)
then the expected root mean squared error is bounded by
\begin{eqnarray*}
\overline{\rm RMS} &\approx&
\frac{1}{2^{m}}\sum_{i=0}^{2^{m}-1}\sqrt{\frac{1}{T}\sum_{t=1}^{T}i^{2}}
\approx
\frac{1}{2}{2^{m}}
\\
{\rm VAR}_{\rm RMS} + \overline{\rm RMS}^{2} &=&
\frac{1}{2^{m}}\sum_{i=0}^{2^{m}-1} \frac{1}{T}\sum_{t=1}^{T}\left|i-A_t\right|^{2}
\approx
\frac{1}{2^{m}}\sum_{i=0}^{2^{m}-1} i^{2}
\approx
\frac{1}{3}({2^{m}})^{2}
\\
{\rm VAR}_{\rm RMS} &\approx&
\frac{1}{3}({2^{m}})^{2} - \frac{1}{4}({2^{m}})^{2}
\\
{\rm SD}&\approx&
\frac{1}{2\sqrt{3}}{2^{m}} = 0.2886751\ 2^{m}
\end{eqnarray*}
%
On the other hand if exhaustive testing is carried out 
and the target values are uniformly spread in the range 
$0 \ldots\ 2^{m}-1$ then 
$\overline{\rm RMS}
\approx \frac{7}{12\sqrt{3}}2^{m}$
and
${\rm SD} \approx
0.23\ 2^{m}$.
\index{Root mean squared (RMS) fitness|)}

\index{Multiplexor benchmark|(}
\section{6~Multiplexor}
\label{sec:6mux}
\begin{figure}
\centerline{\includegraphics{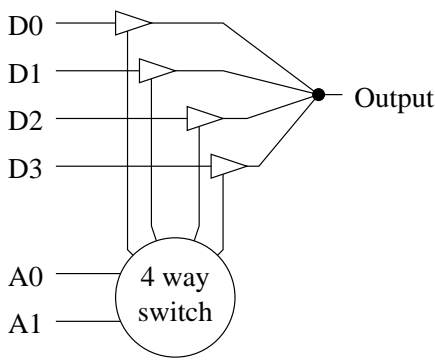}} 
\caption{\label{fig:6mux}
Six way multiplexor.
Only one of four data lines
(D0, D1, D2 and D3) is enabled.
Which one is determined by the two control lines
(A0 and A1).
}
\end{figure}

The six multiplexor problem has often been used as a benchmark
problem.
Briefly the problem is to find a circuit which has two control lines
(giving a total of four possible combinations)
which
are used to switch the output of the circuit to one of the four
input lines,
cf.\
Figure~\ref{fig:6mux}.
\index{Hamming distance fitness|(}
The fitness of a circuit is the number ($0\ldots 64$)
of times the actual output matches the output given by the truth table. 
Note fitness is given by number of bits in common 
between the actual truth table implemented by a program and
a given truth table
(Hamming distance).
\index{Hamming distance fitness|)}

\section{Density of 6~Multiplexor Solutions}
\label{sec:random}

\index{CCNOT!six multiplexor|(}%
\index{Permutation!CCNOT|(}%
\index{CCNOT!all permutations|(}%

\begin{figure}
\centerline{\includegraphics{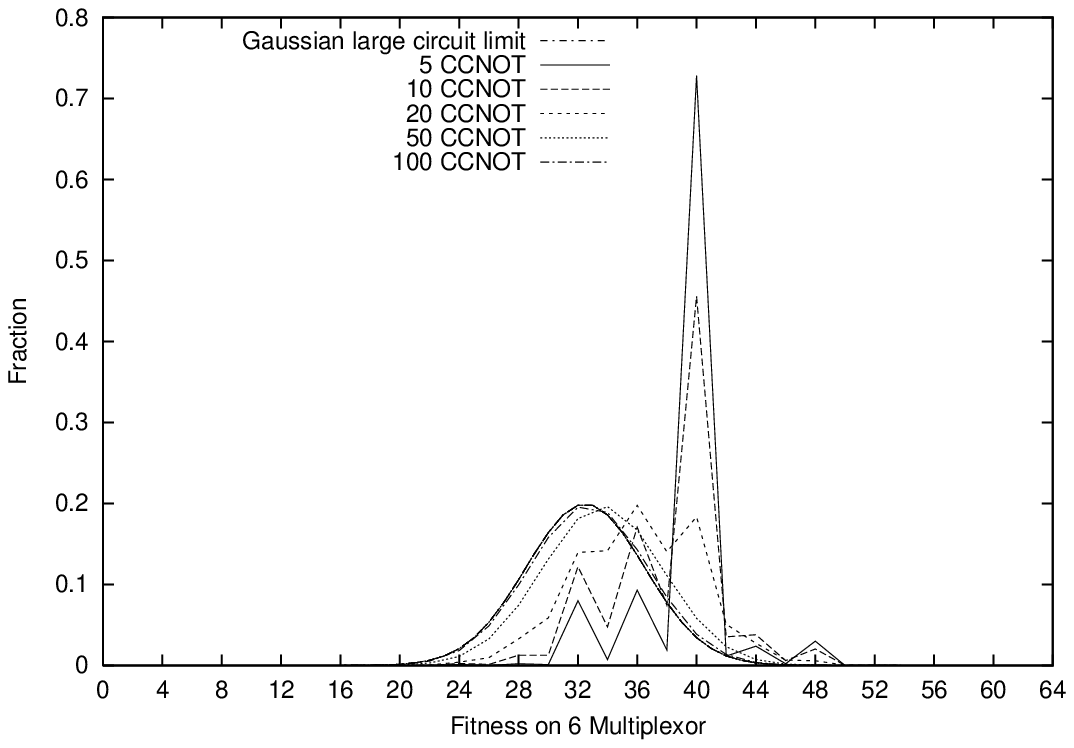}} 
\vspace*{-0.5em}
\caption{\label{fig:CCNOTm6_mux6}
Proportion of 
circuits composed of controlled-controlled-NOT (CCNOT, Toffoli)
gates of each fitness on the 6~multiplexor problem.
Solutions have fitness of~64.
(At least 100~million random circuits tested for each length.)
Since the only wires are those carrying the inputs
(i.e.\
no
additional memory)
odd fitness values cannot be generated.
To simplify the graph these are excluded.
}

\centerline{\includegraphics{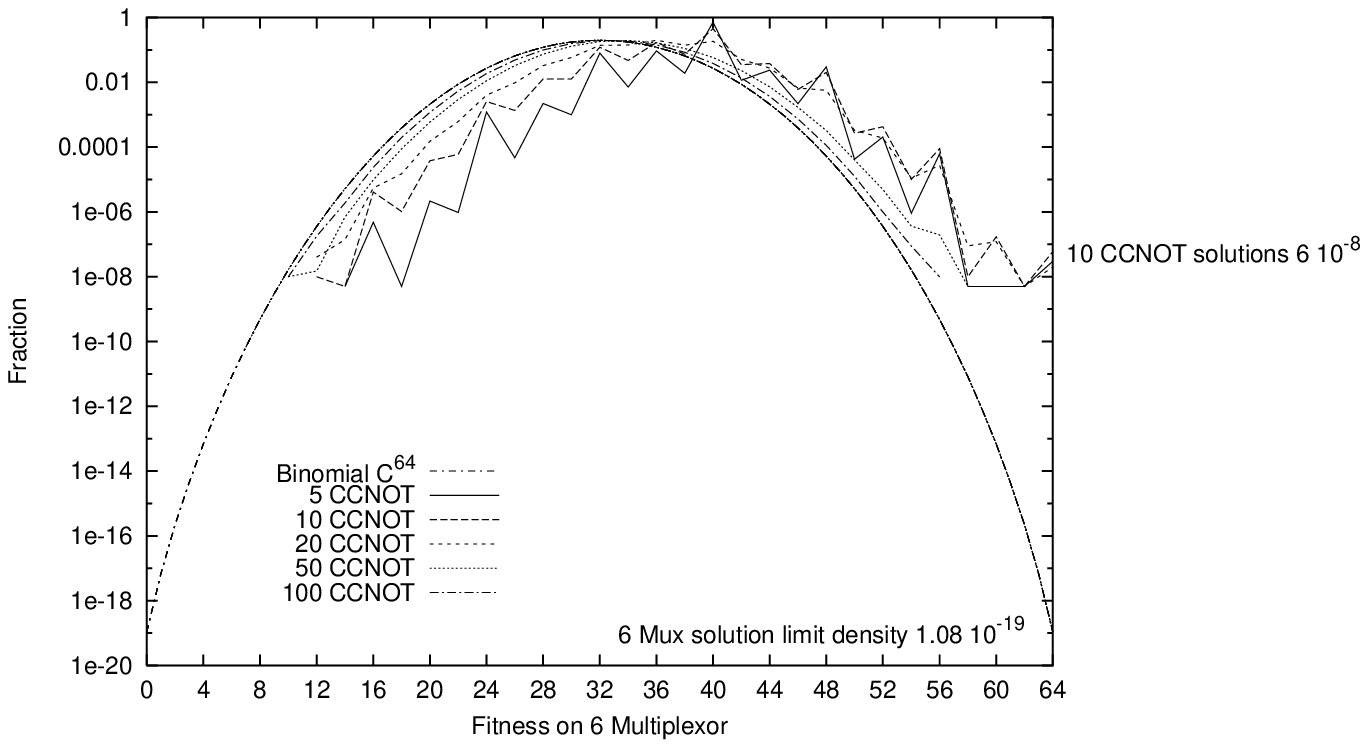}} 
\vspace*{-0.5em}
\caption{\label{fig:CCNOTm6_mux6_log}
Data as Figure \protect\ref{fig:CCNOTm6_mux6}.
In Figure \protect\ref{fig:CCNOTm6_mux6} the large circuit
limit is approximated by
a Normal distribution with mean 32.5.
Here the Binomial distribution approximates the tails
near fitness~0 and~64.
}
\end{figure}

\begin{figure}
\vspace*{-0.3in}
\centerline{\includegraphics{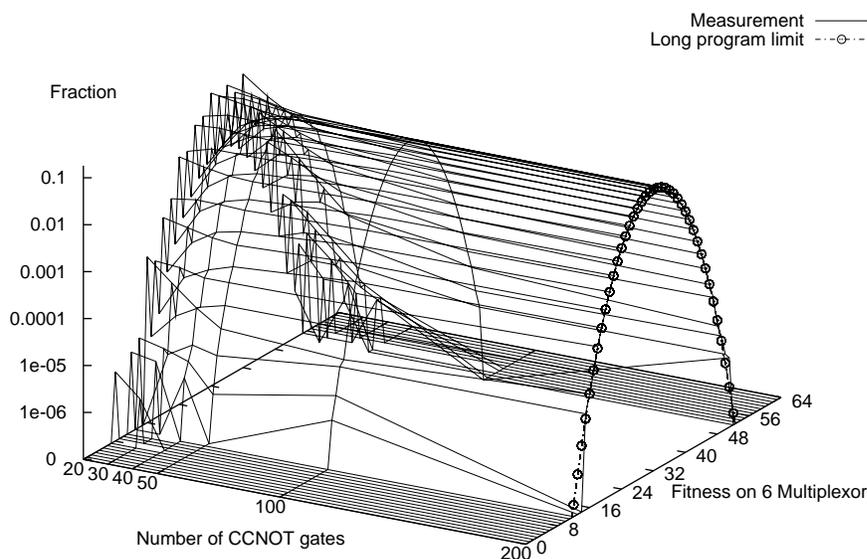}} 
\vspace*{-0.5in}
\caption{\label{fig:CCNOTm7_mux6}
Convergence of 
6~multiplexor 
fitness distribution
as number of CCNOT gates is increased from 20 
towards the large circuit limit
(ringed parabola right hand end).
(At least million random circuits tested for each length.)
One additional memory (garbage) line
ensures all output patterns can be implemented
and in the large circuit limit are equally likely.
I.e.\
the density of solutions is $2^{-64}$.
}
\end{figure}

\begin{figure}
\centerline{\includegraphics{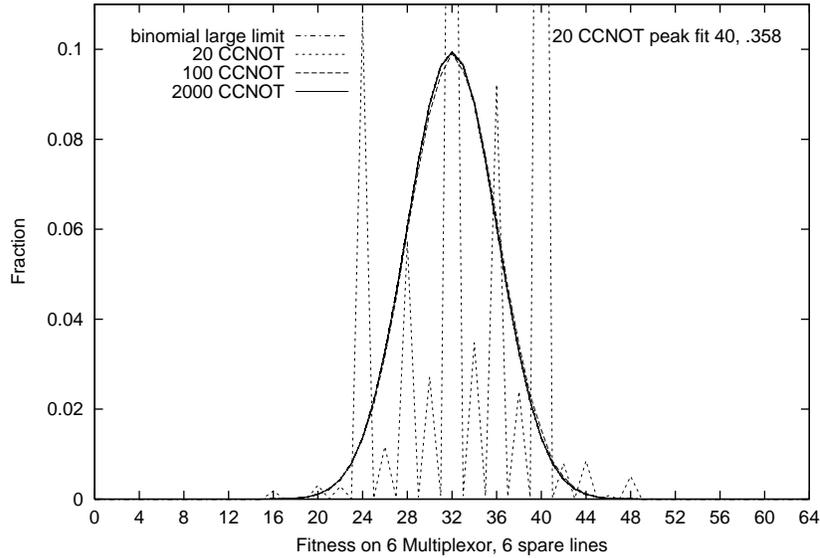}} 
\vspace*{-0.5em}
\caption{\label{fig:CCNOTm12_mux6}
Distribution of fitness on the 6~multiplexor problem 
of circuits of CCNOT gates and with 12 lines.
}
\end{figure}

\begin{figure}
\centerline{\includegraphics{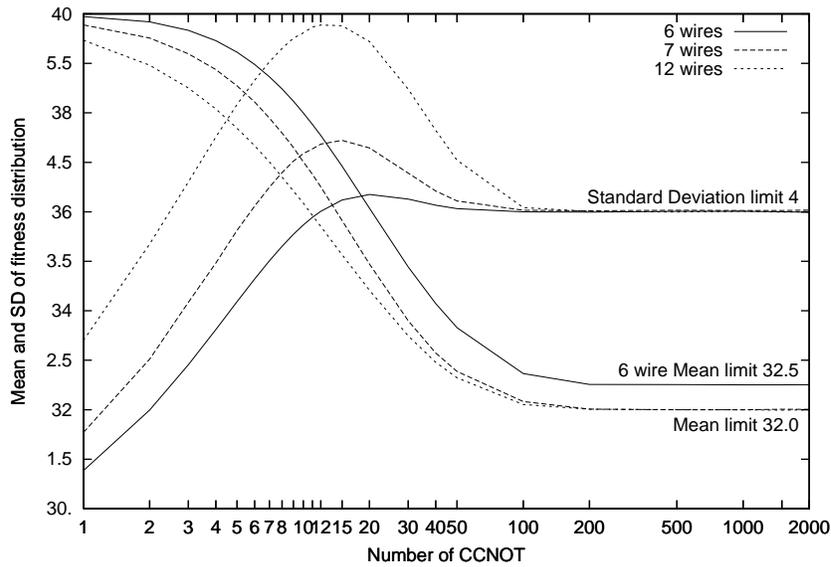}} 
\vspace*{-0.5em}
\caption{\label{fig:CCNOT_mux6_mean}
The fitness distribution of small CCNOT circuits are asymmetric with
mean near 40.
As number of Toffoli gates is increased, both mean and
standard deviation converge to theoretical binomial limits
(32 and 4),
except for circuits without spare wires,
in which case the mean converges to 32.5.
}
\end{figure}

\begin{figure}
\centerline{\includegraphics{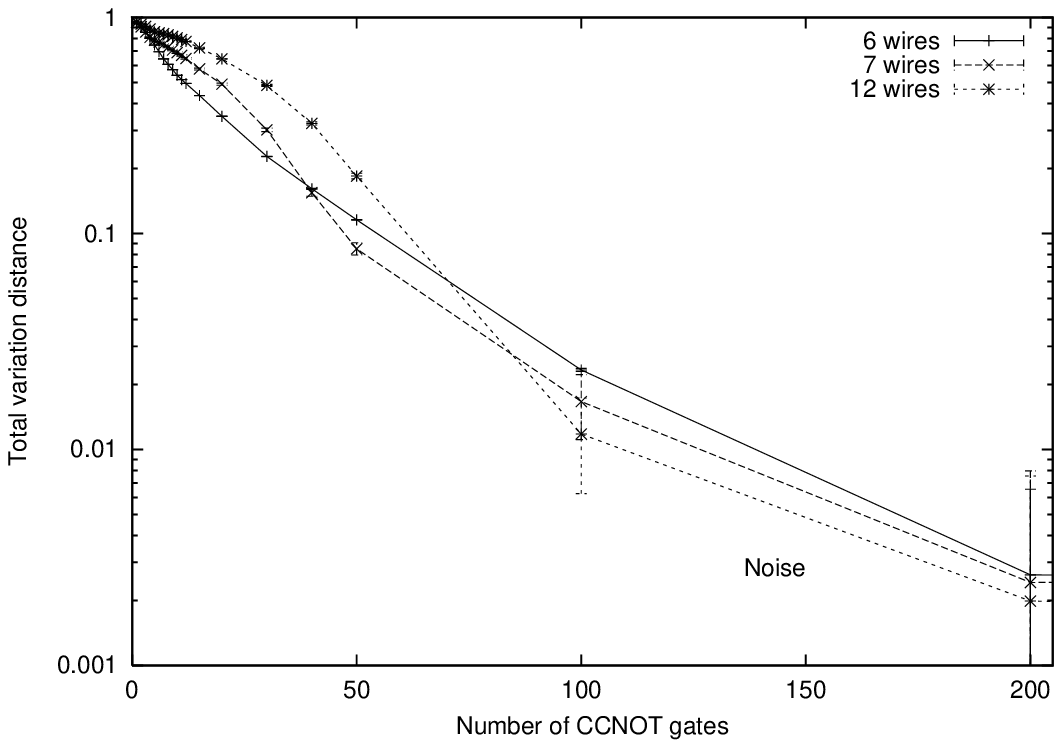}} 
\vspace*{-0.5em}
\caption{\label{fig:CCNOT_mux6_tvd}
Discrepancy between measured distribution of fitness
on CCNOT 6~multiplexor problem
and large circuit limit
(calculated as total variation distance
\protect\cite{rosenthal95convergence}).
Rapid convergence to theoretical limit as program size
increases is shown.
We would expect more spare lines to mean bigger programs are needed 
for convergence.
The plots 
are reminiscent of exponential decay shown for non~reversible programs
\protect\cite{langdon:2002:crlp}
suggesting group theory might lead to results
on the rate of convergence.
}
\end{figure}

We measured%
\footnote{
In all the six multiplexor experiments
we used speed up techniques based on those described in
\cite{poli:1999:aigp3}.%
\index{Speed up!Boolean problems}%
}
the distribution of fitness of randomly chosen CCNOT programs
with 0, 1 and~6 additional wires at a variety of lengths,
cf.~Figures~\ref{fig:CCNOTm6_mux6}--\ref{fig:CCNOT_mux6_64}.
These experiments confirm that there is a limiting
fitness distribution and it is Binomial.
Further,
cf.\
Figure~\ref{fig:CCNOT_mux6_tvd},
the difference between the actual distribution and the limiting
distribution falls rapidly with program length.
Also we do not need many spare wires
(one is sufficient)
to be close to the theoretical
wide circuit limit.
Only without any spare lines
does the limit differ from theory.

\begin{figure}
\index{Multiplexor benchmark!smallest CCNOT solution}
\centerline{
\hfill
\includegraphics{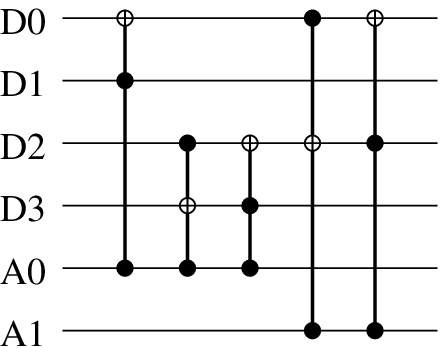}
\hfill
\includegraphics{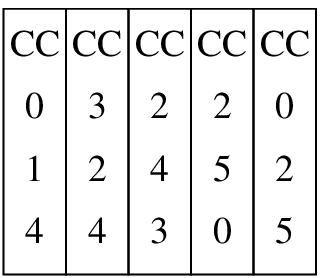}
\hfill} 
\caption{\label{fig:CCNOTp50000m6l5_3}
Example evolved minimal 
circuit (left) of controlled-controlled-NOT (CCNOT, Toffoli)
gates implementing a six way multiplexor.
Genetic representation on right.
Two address lines direct one of the four data inputs to the output.
Circles with crosses indicate controlled wire.
Note there are no additional memory (garbage) lines 
and only five gates are required.
}
\end{figure}

\index{Multiplexor benchmark!smallest CCNOT solution}
There are no CCNOT solutions to the six
multiplexor problem with less than five gates.
One of the smallest solution is shown in
Figure~\ref{fig:CCNOTp50000m6l5_3}.
Notice this does not use any additional wires.
While the multiplexor can be solved by CCNOT without additional storage,
with six lines 
only even fitness scores are possible.
This also means,
even in the limit of long programs,
there is a bias towards higher fitness,
increasing the mean fitness from 
32 to 32.5,
cf.\
Figures~\ref{fig:CCNOTm6_mux6}, \ref{fig:CCNOTm6_mux6_log},
\ref{fig:CCNOT_mux6_mean} and~\ref{fig:CCNOT_mux6_64}.

It is clear that CCNOT has a nice bias for solving the 
six multiplexor problem.
On average small circuits have above average fitness
and in particular 
(at least with six lines)
the chance of solving the problem
is far higher for small programs
than in the limit of large programs.

\index{Permutation!CCNOT|)}
\index{CCNOT!all permutations|)}

\index{Multiplexor benchmark!hill climbing|(}
\index{Hill climbing!six multiplexor|(}
\index{GP!six multiplexor|(}
\index{GP!reversible|(}
\section{Hill Climbing and Evolutionary Solution of the Six~Multiplexor Problem}
\label{sec:mutation}

Section~\ref{sec:proof} tells us how many solutions there are
but not how easy they are to find.
To investigate this
we carried out 
hill climbing  and population based search
on the minimal circuit
and a larger circuit.
Our results are summarised in Table~\ref{tab:solutions}.

\begin{table}[b]
\caption{\label{tab:solutions}%
Number of runs solving the six multiplexor problem
}
\begin{center}
\begin{tabular}{rr|*{2}{r@{/}l}}
\multicolumn{2}{c}{Configuration}&
\multicolumn{2}{c}{Hill climber}& 
\multicolumn{2}{c}{Population}\\ \cline{3-6}
 6 lines &  5 CCNOT & \hspace{1em} 
                      0 &10 &  4&10  \\
12 lines & 20 CCNOT & 1 &10 & 10&10
\end{tabular}
\end{center}
\end{table}

\begin{figure}
\centerline{\includegraphics{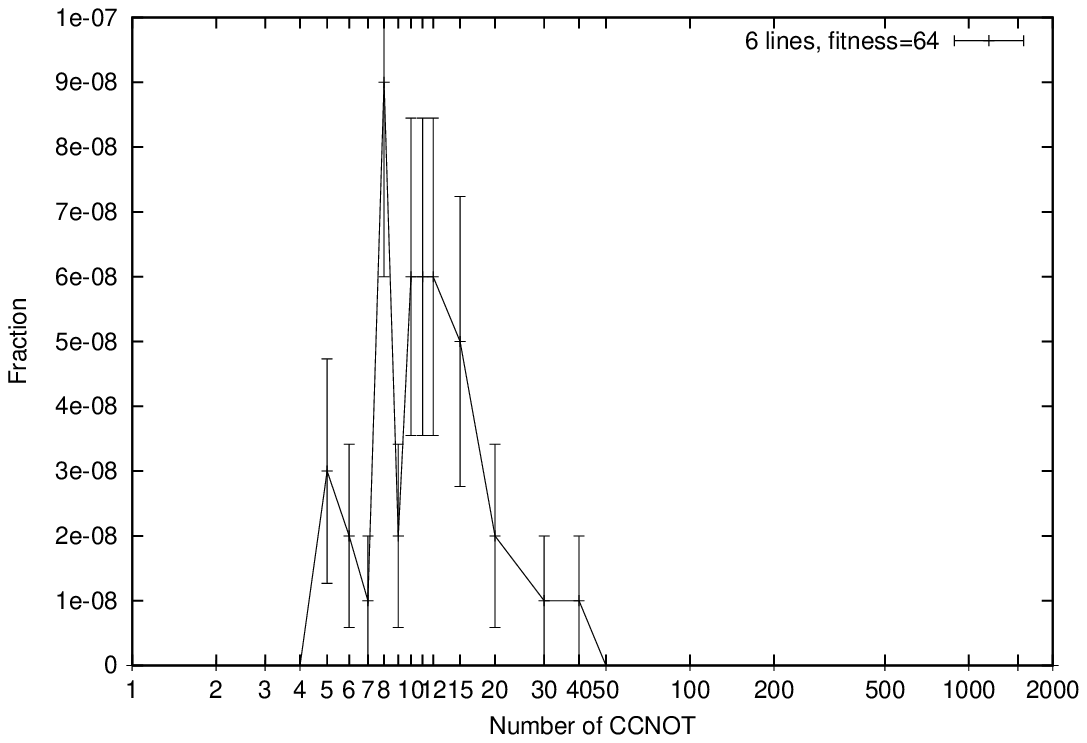}} 
\vspace*{-0.5em}
\caption{\label{fig:CCNOT_mux6_64}
Measured density of solutions
on CCNOT 6~multiplexor problem.
(Based on measurements of at least 100,000,000 random 
circuits of each length.
In one million random samples at each length,
with either one or six spare lines
no solutions were found.)
}
\end{figure}

Firstly we compare random search with these two more sophisticated
search techniques.
Figure~\ref{fig:CCNOT_mux6_64} shows the chance of solving the
six multiplexor by random search of CCNOT circuits.
From the first solutions composed of 5 gates and no spare lines,
the chance rises from about
$3\ 10^{-8}$ to a peak of about 
$10\ 10^{-8}$ at 10 gates,
and then falls towards the theoretical (non~zero) limit of
$5.4\ 10^{-20}$ as the circuit size increases.
(No solutions were found in more than 10,000,000
trials with either one or six spare lines.)

A local hill climber was run ten times.
It mutated exactly one CCNOT gate
of the five in the six wire circuit
and retained the mutant only if its fitness was better.
In seven runs a fitness level of 56 was reached.
%
%
%
%
In two of the seven runs
the hill climber reached fitness~56 significantly%
\footnote{All significance tests in Section~\ref{sec:mutation}
use a 5\% two tailed test.}
faster than
random search could be expected to.
In seven of the remaining runs there is no significant difference between
the hill climber and random search.
(However random search always has a chance of 
solving the problem.
While the hill climber was stuck at local optima and
could never proceed.)
In the remaining run, the hill climber took significantly longer
to reach fitness level~48.

Our hill climber found one solution in ten runs,
when the number of CCNOT gates was increased to 20 and 
six spare lines were added.
Six of the runs attained a score of~56, two runs~48 and the last~52.
None of these nine runs first reached its final fitness level
significantly faster or slower than random search might be expected to.
Except again random search might improve,
whereas
the hill climber was stuck at local optima and
could never proceed.
(Each 20 CCNOT 12~wire program has
$20\times (2 \times (12-2)+(12-3)) = 580 $
or
$20\times 3 \times (12-2) = 600 $
neighbours,
so our hill climber will take on average no more than 
4185 attempts
%
to try them all,
cf.\
The Coupon Collector's problem
\cite[page~284]{feller:1957:ipta}.
All runs had many more trials than this.)
%
%
%
%
%

\index{Evolvability|(}
In contrast the search space turns out to be very friendly
to evolutionary search.
Using the same mutation operator
and a population of 500
(see also Table~\ref{mux.details})
in four out of ten runs
minimal solutions to the six multiplexor were found.
\index{Multiplexor benchmark!smallest CCNOT solution}
(I.e.\
five CCNOT without spare wires).
These solutions took between 20 and 100 generations.
In the remaining six runs fitness level 56 was reached
(in five cases significantly faster than random search).
%
%
%
Introducing six spare wires and extending the circuits from five to 20
CCNOT
makes the problem significantly easier.
Ten out of ten runs (with the same population size etc.)\ 
found solutions.
The solutions were found after 30--220 generations.
This is significantly better than
hill climbing
and population search of the smaller circuit size.
\index{Multiplexor benchmark!hill climbing|)}
\index{Hill climbing!six multiplexor|)}
\index{GP!six multiplexor|)}
\index{GP!reversible|)}
\index{Evolvability|)}

\begin{table} 
\setlength{\temp}{\textwidth}
\settowidth{\tempa}{Functions set:}
\addtolength{\temp}{-\tempa}
\addtolength{\temp}{-2\tabcolsep}
\caption{
Parameters for Multiplexor Problem
}
\label{mux.details}
\begin{tabular}{@{}lp{\temp}@{}}\hline		
Objective: \rule[1ex]{0pt}{6pt}
& Find a reversible function whose output is the same as 
the Boolean 6~multiplexor function
\\ 
Inputs: & D0 D1 D2 D3 A0 A1 (plus either 0 or 6 ``true'' constants)
\\ 
Functions set: & CCNOT
\\ 
Fitness cases: & All the $2^{6}$ combinations of the 6~Boolean
arguments
\\ 
Fitness:       & number of correct answers
\\ 
Selection:     & Tournament group size of 7, non~elitist, generational
\\ 
Pop size:& 500
\\ 
Program size:   & 5 or 20 CCNOT (Toffoli) reversible gates
\\ 
Parameters:    & 100\% mutation
\index{Mutation!reversible gates}%
\index{CCNOT!mutation}%
(Exactly one CCNOT gate is randomly chosen.
One of its three wires
is chosen at random,
and replaced by a randomly chosen, different, but legal, wire.)
\\ 
Termination:   & fitness=64 or 
maximum number of generations G = 500%
\rule[-6pt]{0pt}{6pt}
\\ \hline
\end{tabular}
\end{table}

The effort 
\cite[page~194]{koza:book}
required to find a non-minimal reversible solution
to the six multiplexor 
using a population approach
(87,000) 
is somewhat similar to that 
required by genetic programming to find a non-reversible one
\cite[Table~25.2]{koza:book}.
In other words, using CCNOT with spare lines
has not been shown to be uncompetitive with existing approaches.

\section{Discussion}

It is important to realise the limiting distribution
results hold in general for
reversible computing.
Not just for the six multiplexor or similar problems and 
for any reversible gate,
not just the Toffoli (CCNOT) gate.
An interesting extension would be quantum computing.
Random matrices theory may give a formal bound on the size
of circuits needed to approach the limiting distribution.

\index{Evolvability|(}
The benchmark can be solved with no spare wires.
Indeed
\cite[page~636]{ICALP::Toffoli1980} 
describes a set of gates for which
no more than $m$
spare wires are needed for any finite reversible function.
However
allowing modest increases in the size of solution
by allowing more gates and spare wires
appears to make the fitness landscape more evolvable.
I.e.\
easier for evolutionary search to find solutions.
It is not clear whether
additional gates or wires or both is primarily responsible.
In our example at least,
and we suggest perhaps to other problems,
insisting upon {\em minimal} solutions rather than sufficient
solutions, which may be bigger,
makes the problem unnecessarily hard.
\index{Evolvability|)}

\index{NFL!reversible computing}
\index{No free lunch!see{NFL}}
While NFL 
\cite{ieee-ec:Wolpert+Macready:1997}
applies to reversible computing,
we expect evolutionary search
also to be better than hill climbing and random search
when used with other reversible gates,
\index{Fredkin gate}%
such as the Fredkin gate,
on this and similar problems.

The number of programs or circuits of a given size increases
exponentially with circuit size.
Thus average behaviour across all programs
is dominated by the behaviour of the longest programs.
Almost all of these will behave as the limiting distribution suggest.
Thus considering only the limiting distribution is sufficient to
describe the vast majority of programs.

Where spare wires are included,
for most programs,
the Binomial distribution can be approximated by a Normal distribution
with the same mean and variance.
I.e.\
where fitness is given by a Hamming distance,
the average fitness is $\frac{1}{2}m 2^{n}$
and the variance is $1/2\ 1/2\ m 2^{n}$.
If we normalise fitness to the range $0\ldots 1$,
then the mean becomes 0.5 and the standard deviation is 
$\frac{2^{-\frac{n}{2}}}{2 \sqrt{m}}$.
\index{Hamming distance fitness}%
Even with modest numbers of input and output wires,
the Hamming fitness distribution becomes a needle,
with almost all programs having near average fitness.

In non-trivial problems $n$ and $m$ rapidly become too large to allow
exhaustive testing.
However the limiting distribution still applies.
In the limit
the chance of a random program 
passing
non-exhaustive testing 
is given by the number of bits which are checked.
I.e.\
if $T$ tests are run
and only a $p$ precision answer is needed,
the chance of passing a test case is $2^{-p}$.
The chance of passing all the test cases
is $2^{- T \times p}$.
But note that,
randomly passing the test cases
gives no confidence that the program will generalise.
If an additional independent test is added,
the chance of randomly passing it is only 
$2^{- p}$
\cite{langdon:fogp}.
In contrast general solutions have been evolved
via limited numbers of test cases by
genetic programming
\cite{langdon:book}
suggesting GP has a useful bias for problems of interest.

\section{Conclusions}
\label{conclusions}

\begin{table}
\caption{\label{tab:summary}%
Distribution of Fitness of Large and Wide Reversible Circuits
}
\index{Hamming distance fitness}%
\index{Root mean squared (RMS) fitness}
\begin{center}
\begin{tabular}{|c|ccc|}\hline
\rule[1ex]{0pt}{6pt}
Fitness function & Mean  & Standard Deviation & Perfect Solutions \\ \hline
Hamming 
\rule[1ex]{0pt}{6pt}
& $\frac{1}{2} m 2^{n}$ & $\frac{1}{2} \sqrt{m 2^{n}}$ & $2^{-m 2^{n}}$ \\[1ex]
Normalised Hamming
& $\frac{1}{2}$         & $\frac{1}{2 \sqrt{m 2^{n}}}$ & $2^{-m 2^{n}}$ \\[1ex]
RMS (small T)
& $\frac{1}{2}{2^{m}}$	& $0.29 \ 2^{m}	$	       & $2^{-mT}$ \\[1ex]
RMS (large no.\ tests)
& $0.34\ {2^{m}}$	& $0.23 \ 2^{m}	$	       & $2^{-mT}$ \\[1ex]
\hline
\end{tabular}
\end{center}
\end{table}

As with traditional computing,
as reversible circuits get bigger the distribution of their
functionality converges to a limit.
Therefore their fitness distribution must also tend to a limit.
Table~\ref{tab:summary}
summarises the limit
(with many wires)
for fitness functions based on Hamming distance and
root mean error squared.
In the limit,
every implementable permutation is equally likely.
Note, unlike traditional computing, 
in the limit there is a finite chance of finding a solution.

\index{Evolvability|(}
Experiments on the six multiplexor problem 
have found solutions,
including minimal solutions,
cf.\
Figure~\ref{fig:CCNOTp50000m6l5_3}.
These experiments
suggest
the fitness landscape is amenable to evolutionary search,
particularly if  non-minimal solutions are allowed.
Which in turn suggests the use of variable length evolution.
Performance with CCNOT (Toffoli) gates is similar to that
of genetic programming
with non~reversible programs. 
Simple hill climbing is liable to become trapped at sub~optima,
particularly if constrained to search for minimal solutions.

We suggest that the common emphasis on minimal solutions is misplaced.
These examples provide additional evidence that requiring tiny
solutions hurts evolvability (and other search techniques).
There may only be one smallest program but there are exponentially many
larger solutions.
\index{Evolvability|)}

\index{Multiplexor benchmark|)}
\index{CCNOT!six multiplexor|)}
\index{Reversible computing|)}

\acknowledgments

I would like to thank Tom Westerdale, Ralph Hartley,
Tina Yu, Wolfgang Banzhaf,
Joseph A. Driscoll, Jason Daida, Lee Spector
and Tracy Williams.


\chapbibliography{gp-bibliography,quantum-computing,references}

\end{document}